\documentclass[pra,twocolumn,showpacs,superscriptaddress,byrevtex]{revtex4}

\usepackage{graphicx}
\usepackage{dcolumn}
\usepackage{amsmath}
\usepackage{epsfig}
\usepackage{bm}
\usepackage{amssymb}
\usepackage{amsfonts}

\newcommand{\ket}[1]{\ensuremath{|\,{#1}\,\rangle}}
\newcommand{\bra}[1]{\ensuremath{\langle\,{#1}\,|}}

\newcommand{\itgf}[1]{\ensuremath{\int\!\!d{#1}\,}}

\newcommand{\sinc}{\ensuremath{\mbox{\hspace{1.3pt}sinc}\,}}

\begin{document}


\title{Interference and complementarity for two-photon hybrid entangled states}

\date{\today}

\author{W. A. T. Nogueira}
\email{wallon.nogueira@cefop.udec.cl}
\author{M. Santiba\~{n}ez}
\affiliation{Center for Optics and Photonics, Universidad de Concepci\'{o}n, Casilla 4016, Concepci\'{o}n, Chile}
\affiliation{Departamento de F\'{i}sica, Universidad de Concepci\'{o}n, Casilla 160-C, Concepci\'{o}n, Chile}
\author{S. P\'adua}
\affiliation{Departamento de F\'{\i}sica, Universidade Federal de
Minas Gerais, Caixa Postal 702, Belo~Horizonte,~MG 30123-970,
Brazil}
\author{A. Delgado}
\author{C. Saavedra}
\author{L. Neves}
\author{G. Lima}
\affiliation{Center for Optics and Photonics, Universidad de Concepci\'{o}n, Casilla 4016, Concepci\'{o}n, Chile}
\affiliation{Departamento de F\'{i}sica, Universidad de Concepci\'{o}n, Casilla 160-C, Concepci\'{o}n, Chile}



\begin{abstract}

In this work we generate two-photon hybrid entangled states (HES), where the polarization of one photon is entangled with the transverse spatial degree of freedom of the second photon. The photon pair is created by parametric down-conversion in a polarization-entangled state. A birefringent double-slit couples the polarization and spatial degrees of freedom of these photons and finally, suitable spatial and polarization projections generate the HES. We investigate some interesting aspects of the two-photon hybrid interference, and present this study in the context of the complementarity relation that exists between the visibilities of the one- and two-photon interference patterns.

\end{abstract}

\maketitle


\section{Introduction}

Complementarity of one- and two-particle interference was first proposed by Horne and Zeilinger \cite{Simposio}. They noted that when the two-particle visibility is one, the one-particle visibility observed in either subsystem must be zero and vice-versa. A detailed investigation of this complementarity relation was done by Jaeger \emph{et al.}~\cite{Jaeger1,Jaeger2} for the case of a two-particle beam-splitter based interferometer. They demonstrated that the relation
\begin{eqnarray}
V_{12}^2 + V_{1}^{2} = 1,
\label{complementarity}
\end{eqnarray} holds for any pure bipartite state. Here, $V_{1}$ is the standard one-particle visibility and $V_{12}$ is a two-particle visibility that is obtained from a ``corrected" joint detection probability~\cite{Jaeger1}. This correction is necessary in order to interpret the values of two-particle visibility as a scale, from $0$, for a product state, until $1$, which occurs for a maximally entangled state.

A complementarity relation for the visibilities of the one- and two-photon interference patterns was also derived for a two-photon double-slit
interferometer~\cite{Group5,Horne}. It consists of a source $S$, two double-slits
and two detection screens. In the limit of small source aperture, the two-photon probability amplitude reduces to
$\Psi(1,2)=\cos(k\,\theta^{\prime}\,x_{1})\,\cos (k\,\theta^{\prime} \,x_{2})$, where $\theta^{\prime}$ is the angle that is subtended by the slit
pairs and the detecting planes. In the limit of large source aperture, the two particle probability amplitude reduces to $\Psi(1,2)=\cos
(k\,\theta^{\prime}\,(x_{1} - x_{2}))$ and there is no single photon interference patterns. In this case, the two-photon interference pattern presents ``conditional fringes'', which means that the shape of the fringes depends on the position of both detected photons~\cite{Mandel}. When the source aperture is intermediate in size, both single- and conditional two-photon fringes are present. Horne~\cite{Horne} derived a relation for the one- and the two-photon fringe visibilities that is equal to Eq.~(\ref{complementarity}). This relation was experimentally observed in \cite{Saleh1}. Further experimental investigation of this complementarity relation has been presented recently, where the authors considered distinct classes of spatially entangled two-photon states to test Eq.~(\ref{complementarity}) \cite{Exter2}.

In fact, Eq.~(\ref{complementarity}) is a special case of the expression which fully quantifies the complementarity of bipartite two-level quantum systems~\cite{Bergou1,Bergou2}. It is given by three mutually exclusive quantities that have all the information that can be extracted from the quantum state
\begin{eqnarray}
C^{2} + V_{k}^{2} + P_{k}^{2} = 1,
\label{complementarity2}
\end{eqnarray}
where $k=1,2$. $V_{k}$ is the one-particle visibility and $P_{k}$ is the path predictability, which in the context of a double-slit experiment gives the knowledge that is available about which slit the photon has passed by~\cite{Tessier,Gao,Hosoya,Bergou2,Suter,Huber,Bergou1}. The quantity $C$ is the concurrence~\cite{Wootters}, which is a measurement of the entanglement of the composite system. It corresponds to the maximum value of the two-photon visibility, $V_{12}$, that can be reached. It also represents the information that cannot be extracted if a measurement is performed only on a single particle of the composed system. Because the quantity $C$ cannot be changed by means of a local unitary operations, the quantity $V_{k}^{2} + P_{k}^{2}$ is invariant under this kind of operations. A scheme that allows the measurement of all quantities that are present in Eq.~(\ref{complementarity2}) is showed in Ref.~\cite{Steve}. Complementarity relations involving multi-particle states have been also developed~\cite{Jakob,Tessier,Hosoya,Gao,Suter,Huber,Bergou2}.

Recently, the experimental investigation of hybrid photonic entanglement (HPE), namely the entanglement between two different degrees of freedom (DOF) of composite quantum systems, has been receiving a growing amount of attention~\cite{Zukowsky,ZeilingerBell,BorrQ,Hibr,Barreiro,Tomita,Tittel,Leuchs}. The correlations of the fields generated in the process of spontaneous parametric down conversion (SPDC) \cite{Burnham,Mandel2} has been used for the generation of momentum-polarization entangled states \cite{ZeilingerBell,BorrQ,Hibr,Barreiro}, and also for the generation of time-bin-polarization correlated quantum systems \cite{Tomita,Tittel}. It has also been demonstrated the generation of HPE with continuous variables systems \cite{Leuchs}. The interest in the HPE comes from the fact that they allow more versatility for the quantum communication optical networks. Furthermore, the momentum-polarization hybrid entangled states can be seen, in some cases, as vector-polarization states \cite{Barreiro}, which have many practical potentials \cite{Zhan,Leuchs2}.

In this work, we investigate interference and complementarity aspects for the source of hybrid photonic entanglement (HPE) introduced in Ref.~\cite{Hibr}. In our case, hybrid entangled states (HESs) are prepared in the polarization and transverse spatial degrees of freedom (DOFs) of a photon pair produced in SPDC. One of these photons is sent through a birefringent double-slit (BDS), which discretizes its transverse momentum and couples the photon pair polarization and transverse momentum degrees of freedom. After a suitable choice of spatial and polarization projections, the HES is generated.

For these states we study properties of the two-photon hybrid interference, that naturally arises when the down-converted photons are let to propagate through the free space. The conditional two-photon interference is shown to be dependent only on the angle of the polarization projections performed in one photon, and on the transverse spatial projections implemented onto the second photon, therefore, we refer to it as two-photon hybrid interference. We present this study in the context of the complementarity relation that exists between the visibilities of the one- and two-photon interference patterns. One- and two-photon visibilities are obtained from a set of experimentally produced HESs and their values are used in order to verify Eq.~\ref{complementarity}.

The article is structured as follows. In Sec. II, we first describe the source of transverse momentum-polarization HES used in our experiment. We show how these states can be obtained, and then we give the theoretical expressions for the one- and two-photon interferences, whose hybrid behavior is analyzed. The complementarity relation for the visibilities of these interferences is also discussed, and we show how this relation can properly be verified in an experiment. The theoretical description of this section is fully in accordance with the experiment performed, and thus, the reader may use the setup diagram given in Fig.~\ref{Fig:Setup} to follow the calculations done. The experimental tests of this complementarity relation are presented in Sec. III, and it is followed by the concluding remarks in Sec. IV.

\section{Theory}

\subsection{Brief description of the HES source}

A detailed description of our HES source is given in~\cite{Hibr}. Here we briefly describe the main features of this source. We would like to stress that we have employed a type II SPDC source in the geometry of crossed-cones~\cite{Kwiat95}, instead of two type I crystals used in~\cite{Hibr}. Therefore, our initial two-photon state, after spectral filtering and compensation of longitudinal and transverse walk-off effects, is given by
\begin{equation} \label{Kwiat}
\ket{\Psi}=\frac{1}{\sqrt{2}}(\ket{H}_s \ket{V}_i + e^{j \phi_{pol}} \ket{V}_s \ket{H}_i) \otimes \ket{\Psi_{spa}},
\end{equation} where the state $\ket{H}_l$ ($\ket{V}_l$) represents one photon in the propagation mode $l$ ($l=s,i$ denotes the signal and idler propagation modes, respectively) with horizontal (vertical) polarization. We will assume here that $\phi_{pol}=0$. $\ket{\Psi_{spa}}$ represents the spatial part of the two-photon state generated. If the idler photon is sent through a double-slit, $\ket{\Psi_{spa}}$ becomes a discrete entangled state, as can be seen in Ref.~\cite{Hibr}. If we place a spatial filter on the way of the signal photon, it is made a projection onto the spatial mode $|F\rangle$ defined by the double slit, which results in~\cite{Hibr}
\begin{equation} \label{State}
\ket{\Psi_{spa}} = \left(\frac{\ket{+}_i+
e^{j\phi_{spa}}\ket{-}_i}{\sqrt{2}}\right) \otimes \ket{F}_{s},
\end{equation} where $\ket{F}_s\equiv\frac{\ket{+}_s+ \ket{-}_s}{\sqrt{2}}$. The state $\ket{\pm}_{i}$ is a single photon state defined, up to a global phase factor, as~\cite{QudGen}
\begin{equation} \label{base}
\ket{\pm}_{i} \equiv \sqrt{\frac{a}{\pi}} \itgf{q_{i}}
e^{\mp jq_{i}d/2}\sinc(q_{i}a)\ket{q_{i}}.
\end{equation} The state $\ket{+}_i$ ($\ket{-}_i$) represents the state of the idler photon transmitted by the upper (lower) slit of the double-slit. These states form an orthonormal basis for the Hilbert space of the transmitted photon~\cite{QudGen}. Here, $a$ is the half-width of the slits and $d$ is the center-to-center separation between the two slits. The phase $\phi_{spa}$ can be changed by tilting the double-slit, and we will also consider that $\phi_{spa}=0$.

After preparing the two-photon polarization entanglement, the next step to generate our photonic HES is to couple the polarization and spatial DOFs of the idler photon. This can be achieved by quarter-wave plates (QWP) placed behind each slit of the double-slit, with their fast axes orthogonally oriented, as was shown in Fig.~\ref{Fig:Setup}(b). This birefringent double-slit (BDS) can be seen, up to a phase-shift, as a single-photon two-qubit CNOT gate, where the photon polarization is the control qubit and the transverse momentum distribution, the target qubit~\cite{Hibr}. The action of this CNOT can be summarized as $|H\rangle_i|F\rangle_i \!\Rightarrow \! |H\rangle_i|F\rangle_i$, $|H\rangle_i|A\rangle_i \! \Rightarrow \! |H\rangle_i|A\rangle_i $, $|V\rangle_i|F\rangle_i \! \Rightarrow \! j|V\rangle_i|A\rangle_i $ and $|V\rangle_i|A\rangle_i \!\Rightarrow \! j|V\rangle_i|F\rangle_i$, where the states $\ket{F}_i$ and $\ket{A}_i$ are: $\ket{F}_i\equiv\frac{\ket{+}_i+ \ket{-}_i}{\sqrt{2}}$ and $\ket{A}_i\equiv\frac{\ket{+}_i - \ket{-}_i}{\sqrt{2}}$~\cite{MesSPQb}.

From Eqs.~(\ref{Kwiat}) and (\ref{State}), and taking into account the effect of the BDS, it is straightforward to show that the two-photon state, after the idler transmission through the double slit, can be written as $|\Psi\rangle = (1/\sqrt{2})(|V\rangle_s|HF\rangle_i + j|H\rangle_s|VA\rangle_i)$, where we have omitted the factorable spatial state of the signal photon. This is a two-photon three-qubit GHZ type state~\cite{Group6}, which can be filtered to a hybrid entangled state with a polarization projection on the idler photon. This measurement can be done by placing a polarizer after the BDS, which makes a projective measurement in the idler polarization $|P\rangle_{i}=\alpha|H\rangle+\beta e^{j \phi_P}|V\rangle$. The HES, generated after this projection, will be given by~\cite{Hibr}
\begin{equation} \label{hesgen}
|\Psi^{(P)}\rangle = \alpha |H\rangle_s|F\rangle_i + j\beta e^{-j \phi_P}|V\rangle_s|A\rangle_i,
\end{equation} where we have omitted the idler factorable polarization state, $\alpha$ and $\beta$ are $\geq 0$, and $\alpha^{2} + \beta^{2} = 1$
It is important to note that the amount of entanglement of the generated HES can be tuned by changing the polarization projection of the idler photon~\cite{Hibr}. The concurrence of the state Eq.(\ref{hesgen}) is
\begin{equation}\label{Conc}
C=2 \alpha \beta,
\end{equation} and it is equal to zero for a product state, and one for a hybrid maximally entangled state (HMES).

\subsection{The two-photon hybrid interference}

Next we consider the situation where the signal photon is transmitted through a polarization analyzer, and then is detected by a ``bucket'' detector - an opened avalanche photodiode which registers a photon, but not its transverse position. We also assume that the idler photon  is let to propagate freely after its transmission through the BDS and the polarizer, and then, is detected at a distant plane (where the Fraunhoffer approximation is valid) by a ``point-like'' detector - an avalanche photodiode whose aperture is small when compared to the transverse diffraction pattern formed by the idler down-converted beam.

By assuming also that the photoionization of the idler detector is independent of the idler polarization chosen for the generation of the HES, the probability of joint detection will depend only on the signal polarization analyzer orientation ($\theta_s$) and the idler detector transverse position ($x_i$). This probability, $P_{cc}(\theta_{s},x_{i})$, is therefore given by
\begin{equation} \label{GlauberHib}
P_{cc}(\theta_s,x_i) \propto tr\left(\rho_{\Psi}
\hat{E}^{(-)}_i(x_i)\hat{E}^{(+)}_i(x_i) \otimes \hat{P}_s(\theta_s)
\right),
\end{equation} where $\rho_{\Psi}=\ket{\Psi^{(P)}}\bra{\Psi^{(P)}}$ is the hybrid two-photon density operator [See Eq.~(\ref{hesgen})]. The terms $\hat{E}^{(-)}_i(x_i)$ and $\hat{E}^{(+)}_i(x_i)$, are the negative and positive frequency parts of the electric field operator that represents the spatial evolution of the idler photon from the BDS to the detection plane and at the transverse position $x_i$. The positive frequency part is given, up to an irrelevant phase, by
\cite{Mandel2,QubCar}
\begin{equation} \label{ElectricCamp}
\hat{E}^{(+)}_i(x_i) \propto \int dq \, \hat{a} (q)
\exp \left\{ j \left[ qx_{i} - q^{2}\frac{z}{2k} \right] \right\}
\end{equation} where $k$ represents the wave number of the idler down-converted beam, $z$ is the distance from the double slit to the detector plane and $\hat{a} (q)$ is the destruction operator in the transverse spatial mode $q$. The operator $\hat{P}_s(\theta_s)$, represents the action of the signal polarization analyzer, and it is defined in the usual way, by the projector: $\hat{P}_s(\theta_s)=\ket{\theta_s}\bra{\theta_s}$, where
$\ket{\theta_s} \equiv \cos(\theta_s)\ket{H}_s+e^{j\phi_{s}}\sin(\theta_s)\ket{V}_s$.

Taking into account the completeness relation for the Fock states of light, and the definition of the spatial states $\ket{F}_i$ and $\ket{A}_i$, it is straightforward to show that the joint detection probability, $P_{cc}(\theta_s,x_i)$, may be written as
\begin{widetext}
\begin{eqnarray}\label{CC1}
P_{cc}(\theta_s,x_i) &\propto &|\bra{\theta_s} \bra{vac}\hat{E}^{(+)}_i(x_i)\ket{\Psi}|^2 \nonumber \\
& = & \left| \frac{j\beta e^{-j \phi_P} \cos(\theta_s)}{\sqrt{2}}\left(\bra{vac}\hat{E}^{(+)}_i(x_i)\ket{+}_i-\bra{vac}\hat{E}^{(+)}_i(x_i)\ket{-}_i\right) \right. \nonumber \\ & &\left. + \frac{\alpha \sin(\theta_s)}{\sqrt{2}}\left(\bra{vac}\hat{E}^{(+)}_i(x_i)\ket{+}_i+\bra{vac}\hat{E}^{(+)}_i(x_i)\ket{-}_i\right) \right|^2,
\end{eqnarray}
\end{widetext} where $\ket{vac}$ represents the vacuum state. The detectors quantum efficiencies $\eta$, will be assumed to be $\eta=1$. It is straightforward to show that the matrix elements $\bra{vac}\hat{E}^{(+)}_i(x_i)\ket{\pm}$ are given by
\begin{eqnarray}\label{MaEl}
\displaystyle \bra{vac} \hat{E}^{(+)}_i(x_i)\ket{\pm} & \propto & \displaystyle \exp{\left[j \frac{k(x_{i}\mp d)^2}{2z} \right]} \nonumber \\
& & \displaystyle \times \sinc{\left[ \frac{k a(x_{i}\mp d)}{z}\right]}.
\end{eqnarray}
Considering usual values for the parameters of Eq.~(\ref{MaEl}), such as the ones used in our experiment: $a=40$~$\mu$m, $d=250$~$\mu$m, $k=702$~nm and $z=42$~cm, it is reasonable to assume that $\sinc{\left[ \frac{k a(x_{i}\mp d)}{z}\right]} \sim \sinc{\left[ \frac{k a x_{i}}{z}\right]}$ and also that $\exp{\left[\frac{j k x_{i}^2}{2z} \right]} \sim1$. By using Eq.~(\ref{base}) and Eq.~(\ref{MaEl}) to calculate Eq.~(\ref{CC1}), and considering for simplicity that the signal polarization projection are implemented with $\phi_{s}=0$, we get the following two-photon conditional probability distribution
\begin{widetext}
\begin{eqnarray}\label{CC3}
P_{cc}(\theta_s,x_i) & =
&\left(\frac{2ka}{\pi^2 z}\right)\sinc^2\left(\mathcal{A}x_i\right)\left[\beta^2\cos^2(\theta_s)\sin^2\left(\mathcal{B}x_i\right)
+ \alpha^2 \sin^2(\theta_s)\cos^2\left(\mathcal{B}x_i\right) \right. \nonumber \\ & & \left. + \frac{1}{2}\alpha\beta \sin(2\theta_s)\sin\left(2\mathcal{B}x_i\right)\cos(\phi_{P})\right],
\end{eqnarray}
\end{widetext}
where $\mathcal{A}=(ka/z)$ and $\mathcal{B}=(kd/2z)$. This equation presents many interesting aspects that we are going to analyze.

\begin{figure}[tbh]
\vspace{0.5cm}
\begin{center}
\includegraphics[width=0.5\textwidth]{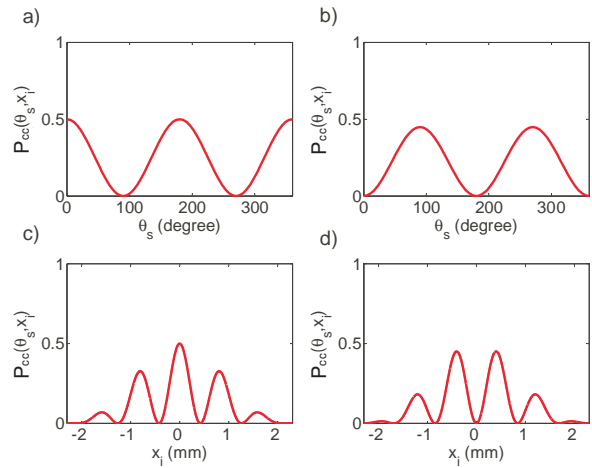}
\end{center}
\vspace{-0.5cm} \caption{(Color online) In (a) and (b), the non-normalized two-photon polarization curves are shown, when the idler detector is fixed at $x_i=\frac{\pi z}{kd}$ and $x_i=0$, respectively, and the signal polarizer analyzer is rotated. In (c) and (d), it is shown the non-normalized two-photon Young's interferences for the cases where the signal polarizer analyzer is set to the vertical and horizontal direction, respectively, and the detector idler scanned transversally. The graphics were made with the values $a=40$~$\mu$m, $d=250$~$\mu$m, $k=702$~nm and $z=42$~cm.}
\label{Fig1}
\end{figure}

\subparagraph{Hybrid behavior -} The two-photon hybrid interference can be measured through completely distinct types of measurements. Whenever the idler detector is fixed and the polarization analyzer is rotated, we will have interference curves that are typical of a polarization-entangled two-photon state~\cite{Kwiat95} (see Fig.~\ref{Fig1}). On the other hand, when the polarization analyzer is fixed and the idler detector is transversally displaced, there will be a conditional fourth-order Young's interference formed~\cite{Mandel}, that has been observed only when spatially correlated photons were transmitted through double-slits~\cite{QubCar,Exter2,Eduardo}.

\subparagraph{The conditionality of the hybrid interference -} In Fig.~\ref{Fig1}(a) and Fig.~\ref{Fig1}(b), the two-photon polarization curves are
plotted in terms of the signal polarizer analyzer angle, when the idler detector position is fixed at $x_i=\frac{\pi z}{kd}$ and $x_i=0$, respectively. When $x_i=\frac{\pi z}{kd}$, the interferences fringes will be governed by the first term of Eq.~(\ref{CC3}), while when $x_i=0$, it will be governed by the second term. One can clearly see the conditionality of the two-photon polarization interference. The two-photon conditional Young's interferences are shown in Fig.~\ref{Fig1}(c) and Fig.~\ref{Fig1}(d), for the cases where the signal polarization analyzer is set to the vertical and horizontal direction, respectively, and the idler detector is scanned transversally. For plotting these curves, we considered the values of $k$, $z$, $a$ and $d$ of the experimental setup used.

\subparagraph{The conditionality of $V_{12}$ -} The visibility of the two-photon hybrid interference can be defined as
\begin{equation} \label{vh}
V_{12}\equiv\frac{[P_{cc}(\theta_s,x_i)]_{max}-[P_{cc}(\theta_s,x_i)]_{min}}{[P_{cc}(\theta_s,x_i)]_{max}+[P_{cc}(\theta_s,x_i)]_{min}}.
\end{equation} As is also the case with polarization-only or spatially-only entangled photons, the hybrid two-photon visibility is dependent on the measurement bases considered, even for hybrid maximally entangled state (HMES). For example, in the case of HMES, the two-photon Young's interference
will disappear when the idler detector is scanned with the signal polarization analyzer oriented at $45^{\circ}$ \cite{Note}. This effect is related with the wave-particle duality of the photon transmitted by the BDS.

\subparagraph{Applications -} As it is well summarized in \cite{Cai}, when there is some prior knowledge about the quantum state, it is possible to use the two-photon interference to determine their amount of entanglement \cite{Wootters}. There have been several works which used this technique for measuring distinct types of quantum correlations \cite{Exter2,QubCar,Franca,Zubairy}. The hybrid two-photon interference, can be used in this way to quantify the hybrid entanglement. As we mentioned before, the concurrence for the HES of Eq.~(\ref{hesgen}) is given in terms of $\alpha$ and $\beta$, which can be determined from the measurements described in Fig.~\ref{Fig1}. The advantage is that one can choose which type of measurement, the two-photon polarization curve or the two-photon Young's interference, is more suitable for the experimental entanglement determination.

\subsection{One-photon interferences}

There are two distinct types of one-photon interference that may be observed, one for each DOF involved in the photonic entanglement. When the counts of the idler detector are registered as a function of its transverse displacement, there will be a Young's second-order interference formed. When the signal detection is registered as a function of the orientation of its polarization analyzer, a sinusoidal curve will be formed.

\subsubsection{The one-photon spatial interference} The probability of detecting the idler photon is proportional to the spatial second-order correlation function, defined as $tr\left(\rho_{i}\hat{E}^{(-)}_i(x_i)\hat{E}^{(+)}_i(x_i)\right)$. $\rho_i$ is the density operator representing the idler photon and it can be obtained by tracing out the polarization of the signal photon from the HES given by Eq.~(\ref{hesgen}). Considering the approximations for the elements of matrix $\bra{vac} \hat{E}^{(+)}_i(x_i)\ket{\pm}$ exposed before, the probability of single detection for the idler detector is given by
\begin{widetext}
\begin{eqnarray}\label{C1s1}
P_{1s}(x_i) & \propto & |\alpha|^2 |\bra{vac}\hat{E}^{(+)}_i(x_i)\ket{F}_i|^2 + |\beta|^2|\bra{vac}\hat{E}^{(+)}_i(x_i)\ket{A}_i|^2 \nonumber \\
& = & \left(\frac{2ka}{\pi z}\right) \frac{1}{2} \sinc^2\left(\mathcal{A}x_i\right)\left[1 + (\alpha^2-\beta^2)\cos\left(\mathcal{B}x_i\right)\right].
\end{eqnarray}
\end{widetext}
We can see that the one-photon spatial interference has a visibility
\begin{equation}\label{v1s}
 V_{1s}=|\alpha^2-\beta^2|,
\end{equation} that is $0$ when the down-converted photons are in hybrid maximally entangled sate and $1$ for a product state. We note that $ V_{1s}$ may also be written as $V_{1s}=\sqrt{1-C^2}$.

The experimental measurement of $P_{1s}(x_i)$ can be easily performed. The trace over the signal polarization is done by removing the polarization analyzer from its propagation path, and the coincidences counts in this case, will map $P_{1s}(x_i)$ while the idler detector is scanned. As it is explained in \cite{Exter2,QubCar}, it is important to understand that the correct measurement of $P_{1s}(x_i)$ must be done using the coincidence counts recorded and not the single counts, since the latter is also formed of photons that do not belong to the two-photon HES of Eq.~(\ref{hesgen}).

\subsubsection{The one-photon polarization interference} The probability of detecting the signal photon is proportional to $tr(\rho_s \hat{P}_s(\theta_s))$. $\rho_s$ is the reduced density operator that represents the signal photon and it is obtained after tracing out the spatial content of the idler photon from the two-photon HES of Eq.~(\ref{hesgen}). So, we have that
\begin{eqnarray} \label{C1p1}
P_{1p}(\theta_s)& \propto &|\alpha|^2|\bra{\theta_s}H\rangle|^2+|\beta|^2|\bra{\theta_s}V\rangle|^2 \nonumber \\
& = & \left(\frac{1}{2\pi}\right)[1 + (\beta^2-\alpha^2) \cos(2\theta_s)],
\end{eqnarray} which has the same visibility of the one-photon spatial interference.

We also note that $P_{1p}(\theta_s)$ can easily be measured. The operation of tracing out the information of the momentum distribution of the idler photon is done by opening the idler detector, which now becomes a ``bucket'' detector, in the sense discussed before. The coincidence counts will then map the $P_{1p}(\theta_s)$ curve, while the polarization analyzer before the signal detector is rotated.

\subsection{Testing the complementarity relation}

As it is discussed in \cite{Jaeger1,Jaeger2}, the definition of the visibility of the two-photon interference pattern given by Eq.~(\ref{vh}) fails to capture the intended sense of two-particle interference, since it yields $V_{12}=1$ even if the two-photon HES is a product state. So, to properly study the one- and two-photon complementarity relation, we adopt the correction proposed by Jaeger \emph{et al.},~\cite{Jaeger1,Jaeger2}, where the presence of a correction factor for the joint probability, allows one to have the corrected two-photon visibility equal to one for a maximally entangled state, and zero for a product state. However, before continuing we would like to emphasize that the consideration of joint probability distribution in the form that was given in Eq.~(\ref{CC3}), is completely relevant since it is the only two-photon distribution that indeed can be measured directly in the laboratory.

The corrected joint detection probability, $\bar{P}_{cc}(\theta_s,x_i)$, is, in accordance with the idea of a correction factor exposed in references~\cite{Jaeger1,Jaeger2}, given by
\begin{eqnarray} \label{CCcorrec}
\bar{P}_{cc}(\theta_s,x_i) & = & P_{cc}(\theta_s,x_i)\!-\!P_{1s}(x_i)P_{1p}(\theta_s) \nonumber \\
& & + \left(\frac{2ka}{\pi^2 z}\right)\frac{1}{4}\sinc^2(\mathcal{A}x_i),
\end{eqnarray} where the extra factor, $\sinc^2(\mathcal{A}x_i)$, accounts for propagation and diffraction effects that appear after the idler photon is transmitted through the BDS. Even though this probability cannot be measured directly, it can be calculated from the experimental results of $P_{cc}(\theta_s,x_i)$, $P_{1s}(x_i)$ and $P_{1p}(\theta_s)$, as it has been done in \cite{Saleh1}, for spatially correlated photons.

By substituting Eqs.~(\ref{CC3}), (\ref{C1s1}) and (\ref{C1p1}) into (\ref{CCcorrec}) one obtains that
\begin{widetext}
\begin{eqnarray} \label{CCcorrec2}
\bar{P}_{cc}(\theta_s,x_i)
& = & \left(\frac{ka}{2\pi^2 z}\right)\sinc^2 \left( \mathcal{A}x_i\right)\{ 1 + C \sin(2\theta_{s}) \sin\left(2\mathcal{B}x_{i}\right)
\cos(\phi_{P}) + C^{2} \cos(2\theta_{s}) \cos(2\mathcal{B}x_{i})\},
\end{eqnarray}
\end{widetext}
which has the same structure of the Eq. (82b) of reference~\cite{Jaeger2}.

Thus, the corrected two-photon visibility becomes
\begin{equation} \label{vhcorrec}
V_{12}\equiv\frac{[\bar{P}_{cc}(\theta_s,x_i)]_{max}-[\bar{P}_{cc}(\theta_s,x_i)]_{min}}
{[\bar{P}_{cc}(\theta_s,x_i)]_{max}+[\bar{P}_{cc}(\theta_s,x_i)]_{min}},
\end{equation} and what is interesting to note is that the complementarity relation given by Eq.~(\ref{complementarity}), can be tested considering four distinct types of measurements: (i) One can choose to measure the one-photon spatial interference and compare its visibility with the two-photon Young's interference corrected visibility or, (ii) compare it with the corrected visibility of the two-photon polarization curve. One can also choose to compare the visibility of the one-photon polarization curve with the corrected visibilities of the (iii) two-photon polarization curve, or (iv) the two-photon spatial curve.

As we mentioned before, the visibility of the two-photon interference depends on the chosen measurement basis. In fact, its value can vary from $C^2 \leq V_{12} \leq C$ \cite{Jaeger2}. Here, we consider the measurement of the conditional polarization curve while the idler detector is fixed at $x_i=0$, and the measurement of the conditional Young's interference when the signal polarizer analyzer is fixed at the vertical direction, $\theta_s=\pi/2$. For these measurements, it is possible to derive simple expressions for the corrected two-photon interferences, and they allow one to clearly see the relation between $V_{12}$ and the concurrence $C$ of the HES. When the signal polarization analyzer is fixed at the vertical direction, the corrected two-photon spatial interference of Eq.~(\ref{CCcorrec}) may be written as
\begin{eqnarray} \label{CCcorrecSpa}
\bar{P}_{cc}(\pi/2,x_i)&=&\left( \frac{ka}{2\pi^2 z} \right)\sinc^2\left(\mathcal{A}x_i\right) \nonumber \\
& & \times \left[1 + C^2 \cos\left(2 \mathcal{B}x_i\right)\right],
\end{eqnarray} which has a visibility $V_{12} = C^2.$ When the idler detector is fixed at $x_i=0$, we will have $\bar{P}_{cc}(\theta_s,x_i)$ in terms of the angle of the signal polarization analyzer given by
\begin{eqnarray} \label{CCcorrecPol}
\bar{P}_{cc}(\theta_s,0)=\left( \frac{ka}{2\pi^2 z} \right)\left[1 - C^2 \cos(2 \theta_s)\right],
\end{eqnarray} that, of course, has the same visibility $V_{12} = C^2$.

\section{Experiment}

\subsection{Experimental setup} 
\begin{figure*}[tbh]
\vspace{0cm}
\begin{center}
\rotatebox{0}{\includegraphics[width=0.8\textwidth]{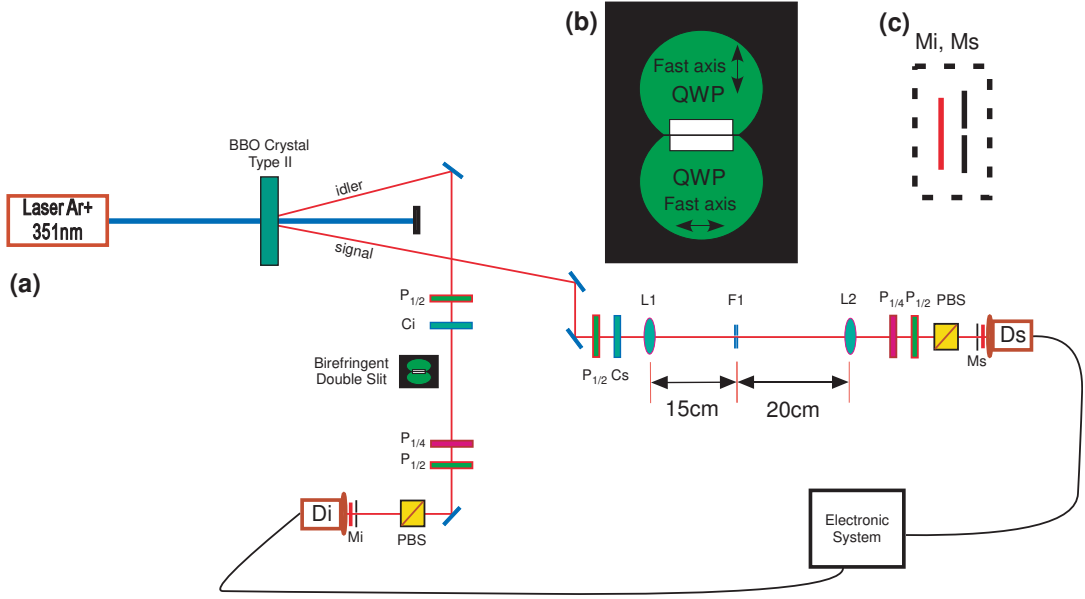}}
\end{center}
\vspace{0cm} \caption{(Color online) Experimental setup. (a) A single-mode Ar$^{+}$-ion laser with wavelength of $351.1$\,nm passes through a BBO type II crystal of $5$\,mm width. In both arms there are half-wave plates (\textbf{P$_{1/2}$}), that rotate the polarization by $90^{\circ}$, followed by a BBO compensating crystal of $2.5$\,mm width (\textbf{Ci} and \textbf{Cs}). Also, half-wave plates, quarter-wave plates, and polarizing beam splitters are placed before detectors in order to make projection in the desired polarization basis. The BDS is placed in the idler arm. In the signal arm there is a spatial filter made by two lens (\textbf{L$1$} and \textbf{L$2$} with $15$\,cm and $20$\,cm of focal distances, respectively) and a horizontal slit, \textbf{F$1$}, with $50$\,$\mu$m width. \textbf{Di} and \textbf{Ds} are the idler and signal single-photon detectors, respectively. (b) Sketch of the BDS: two quarter wave plates with their fast axes in perpendicular directions; one in front of the upper slit and the other in front of the bottom slit. (c) \textbf{Mi} and \textbf{Ms} represent the spatial and frequency filters that are placed before \textbf{Di} and \textbf{Ds}, respectively. They are composed of an iris of controllable diameter followed by an interference filter centered at $702$\,nm with $1$\,nm FWHM.}
\label{Fig:Setup}
\end{figure*}

The experimental setup considered is outlined in Fig.~\ref{Fig:Setup}(a). A single-mode collimated Ar$^{+}$-ion laser operating at $351.1$\,nm, with an average power of $71$\,mW and in a TEM$_{00}$ mode with transverse profile of $\sim2$\,mm FWHM, is sent through a $5$-mm-long $\beta$-barium borate crystal (BBO) cut for type-II SPDC. Degenerated down-converted photons of $702$\,nm are selected using interference filters that have small bandwidths ($1$\,nm FWHM) and are mounted in front of the idler and signal detectors, which will be referred as \textbf{Di} and \textbf{Ds}, respectively, from now on. In order to prepare the $|\Psi^{+}_{pol}\rangle$ polarization state, a half-wave plate and BBO crystals of $2.5$\,mm thickness are placed on each down-converted arm to compensate longitudinal and transverse walk-off effects~\cite{Kwiat95}. At idler path, a BDS is placed at a distance of $40$\,cm from the crystal, followed by a polarization analyzer, which is composed of a QWP, a HWP and a PBS. The BDS is sketched in Fig.~\ref{Fig:Setup}(b). The slit width is $2a=80$\,$\mu$m and their center-to-center separation is $d=250$\,$\mu$m. Upon its transmission through this system, the idler photon propagates freely through $42$~cm, until it reaches the single-photon detector \textbf{Di}, which is mounted on a translation stage that allows its displacement in the $x$ transverse direction.

In the signal path, the spatial mode of the photon is defined by a spatial filter placed after the compensating crystal. It is composed by two lenses and a horizontal slit of $50$\,$\mu$m width, that is placed at the focal plane of both lenses. The first and second lenses have focal distances of $150$\,mm and $200$\,mm, respectively. After the spatial filter, a polarization analyzer (QWP, HWP and PBS) allows for polarization projection in any basis. The detector \textbf{Ds} is located at $25$\,cm from the second lens of the spatial filter. The detectors \textbf{Di} and \textbf{Ds} are connected to a circuit used to record single and coincidence counts.

\begin{figure}[tbh]
\vspace{-0.5cm}
\begin{center}
\includegraphics[width=0.5\textwidth]{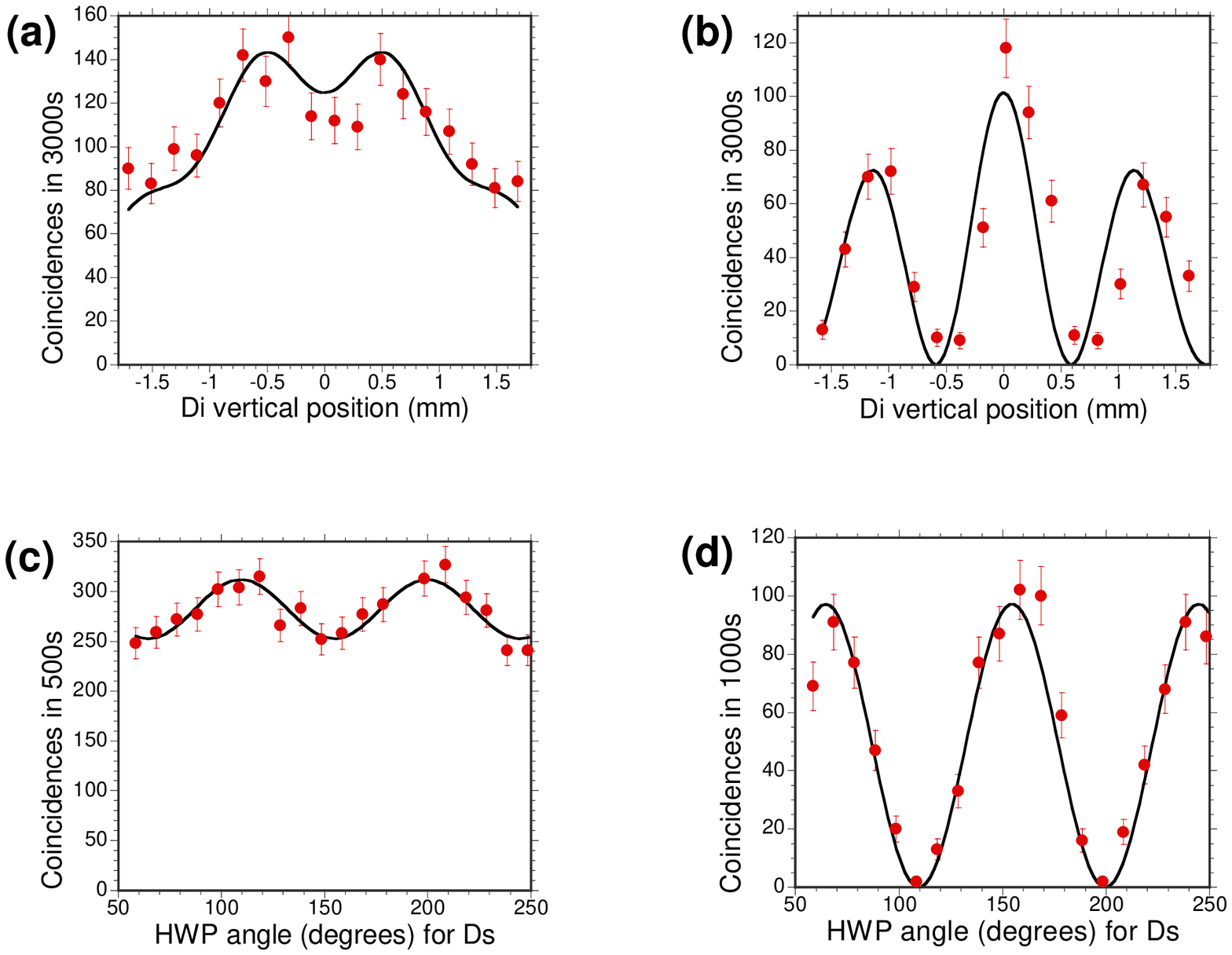}
\end{center}
\vspace{-6cm}\caption{(Color online) Experimental results for HMES. (a) One-photon and (b) two-photon spatial interference patterns. (c) One-photon and (d) two-photon polarization interference patterns. In all figures, vertical error bars represent square-root of the measurements. One can see that there is a good agreement between the theoretical predictions (solid lines) and the experimental results (circles).}
\label{Fig:Med1}
\end{figure}
\vspace{0cm}

\begin{figure*}[tbh]
\vspace{0.0cm}
\begin{center}
\centering \rotatebox{-90}{\includegraphics[width=0.7\textwidth]{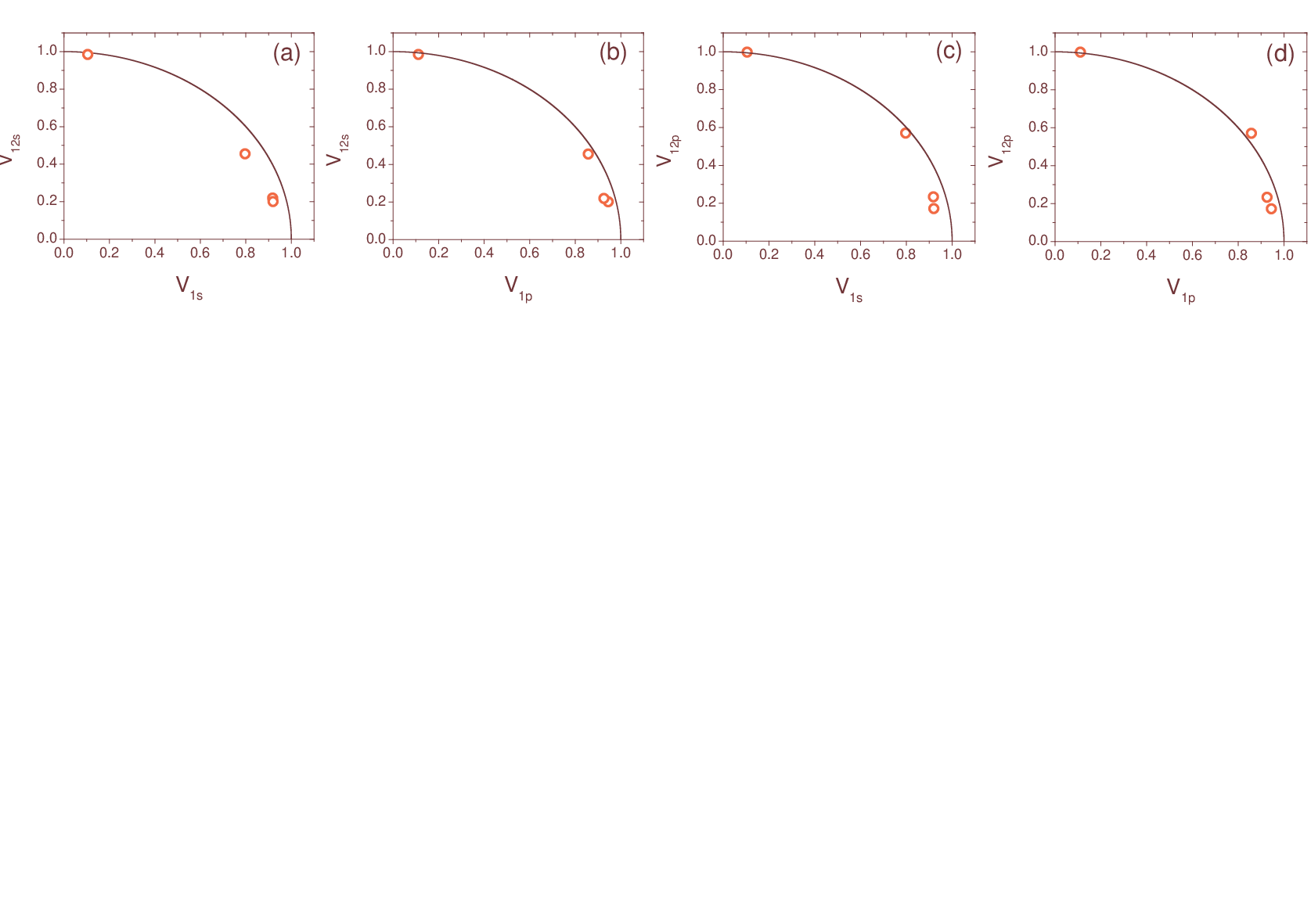}}
\end{center}
\vspace{-8.5cm}
\caption{(Color online) Obtained two-photon visibilities versus the one-photon visibilities. Each circle corresponds to a distinct HES. In (a) it is shown the two-photon spatial versus the one-photon spatial visibilities. Figure (b) shows the two-photon spatial versus the one-photon polarization visibilities. Figure (c) shows the two-photon polarization versus the one-photon spatial visibilities. Figure (d) shows the two-photon polarization versus the one-photon polarization visibilities. The two-photon visibilities for HMES are the greatest (circles on the left in the graphics). The two-photon visibilities for the product states are the smallest (circles on the right in the graphics). The solid line corresponds to the complementarity relation given by Eq.~\ref{complementarity}.} \label{Fig:exCompl}
\end{figure*}
\vspace{3cm}

\subsection{State preparation}

The first step in order to generate different HESs is to ensure that the polarization-entangled-state generated [see Sec. II(A)] has a high degree of  entanglement and purity. In order to verify this point we have measured interference curves in the $+45^{\circ}/-45^{\circ}$ polarization basis, as described in~\cite{Kwiat95}, and observed visibility that reached $0.95$. This has been done before placing the BDS on the idler arm and the spatial filter on the signal arm. Besides, we made the tomographic reconstruction of the two-photon polarization state \cite{James} and observed a purity of $0.92\pm0.02$ ($Tr(\rho^2)$) and a fidelity \cite{Jozsa} of $0.95\pm0.03$ with the state $|\Psi^{+}_{pol}\rangle=(1/\sqrt{2})(|H\rangle_{i}|V\rangle_{s}+|V\rangle_{i}|H\rangle_{s})$.

For testing the complementarity relation, we generated HESs with different degrees of entanglement by means of a suitable choice of the polarization projection implemented in the idler photon, as it was described in Sec. II(A). The first state was a hybrid maximum entangled state, generated with $\alpha=\beta=\frac{1}{\sqrt{2}}$ and $\phi_P=\pi/2$, which corresponds to project the idler photon onto the left-circular polarization. This state is given by
\begin{equation}
|\Psi^{(P)}\rangle = \frac{1}{\sqrt{2}}(|V\rangle_{s}|F\rangle_{i}+|H\rangle_{s}|A\rangle_{i}).
\end{equation}
The other three prepared states are generated by considering distinct linear polarization projections in the idler arm. They have the form
\begin{equation}\label{xi}
\!\!|\Psi^{(P)}\rangle = \cos (2\xi)|V\rangle_{s}|F\rangle_{i} + \sqrt{\cos^2 (2\xi)-1}|H\rangle_{s}|A\rangle_{i},
\end{equation}
and in our case, we adopted $\xi=0^{\circ}$, $\xi=5^{\circ}$ and $\xi=10^{\circ}$.

\subsection{Results}

We measured four types of interference curves for each state: one- and two-photon polarization curves and one- and two-photon spatial interferences from which we obtain the visibilities: $V_{1p}$, $V_{12p}$, $V_{1s}$, and $V_{12s}$, respectively. For these measurements, it was used a $1.3$\,mm diameter iris in front of \textbf{Ds}. The aperture in front of the detector \textbf{Di} was modified depending on the measurement that was performed.

For the measurement of the two-photon polarization curves, a single slit of $100$\,$\mu$m$\times$$5$\,mm ($xy$) was placed in front of detector \textbf{Di}, which was fixed at $x_i=0$, corresponding to spatial projection of the idler photon onto $|F\rangle_i$~\cite{Hibr,MesSPQb}. In the case of one-photon polarization curves, a $2.0$\,mm diameter iris was placed before detector \textbf{Di}. This action turns \textbf{Di} into a ``bucket" detector, which detects a photon without registering its transverse position. This is necessary to trace over the spatial DOF, which means that \textbf{Di} must be unable to distinguish between $|F\rangle_{i}$ and $|A\rangle_{i}$~\cite{QubCar}. The one- and two-photon polarization curves were obtained by measuring the coincidence rate in terms of the HWP angle at the signal arm. Coincidence curves were recorded for each of the four prepared HES.

For the measurements of one- and two-photon spatial patterns, a $100$\,$\mu$m$\times$$5$\,mm slit was placed in front of \textbf{Di}. In the case of one-photon spatial patterns, the signal polarization analyzer in front of detector \textbf{Ds} was removed to perform a trace over the polarization DOF. In the case of the two-photon spatial patterns, the polarization analyzer was fixed at the vertical direction ($\theta_{s}=\pi/2$), determining the conditional measurement that we desired to perform. All spatial curves were obtained by recording the coincidence counts as a function of the \textbf{Di} transverse position ($x$).

Figure \ref{Fig:Med1} shows the measurements based on the procedure described above for the case of the HMES. The graphics of \ref{Fig:Med1}(a) and \ref{Fig:Med1}(b) show the observed one- and two-photon spatial interference patterns, respectively. The one-photon curve was fitted according to Eq.~(\ref{C1s1}). From this curve, we obtained the one-photon spatial visibility, $V_{1s}$, which was, together with the amplitude, a free parameter in the fit. The conditional spatial curve was fitted with Eq.~(\ref{CC3}) by choosing $\theta_{s}=\pi/2$ and leaving the amplitude as a free parameter. In both fits, the vertical error bars are proportional to the square roots of the measured coincidences.

These curves, properly normalized with the factors present in Eqs.~(\ref{CC3}) and~(\ref{C1s1}), were used in the calculation of the two-photon corrected curves. The same procedure was used in order to make the fits of Figs.~\ref{Fig:Med1}(c) and~\ref{Fig:Med1}(d) for the polarization DOF, using the respective equation, i.e., Eq.~(\ref{C1p1}) for the Fig.~\ref{Fig:Med1}(c) and Eq.~(\ref{CC3}), with the choice $x_{i}=0$, for the Fig.~\ref{Fig:Med1}(d).

Based on the curves obtained for the one- and two-photon patterns, it was possible to obtain the corrected coincidence curves for all the four prepared HES. Figure~\ref{Fig:exCompl} summarizes the experimental results obtained for these states, where it is plotted the two-photon vs one-photon visibilities. From left to right, the experimental points (red circles) correspond to HMES and the non-maximally HESs with $\xi$ equal to $10^\circ$, $5^\circ$, and $0^\circ$ (product state), given in Eq.~(\ref{xi}).

The agreement of experimental data and the ideal complementarity relation (solid curve) given in Eq.~(\ref{complementarity}) is good. The discrepancies between theory and experiment can be attributed mainly to the non-perfect polarization entangled state initially prepared, to the non-perfect coupling in the BDS due to the misalignment of the QWP, and also to the polarization projection to generate the HES. Finally there is another source of error in the process of spatial tracing when we measured the one-photon polarization curves. As it was noted above, the iris in front of the detector \textbf{Di} must be sufficiently large in order to not distinguish between $|F\rangle_{i}$ and $|A\rangle_{i}$.

\section{Conclusion}

In this work we extensively analyzed the properties of the two-photon hybrid interference patterns, that naturally arises when the HES are let to propagate through the free space, and presented it in the context of the complementarity relation that exists between the one- and two-photon visibilities. We theoretically obtained the expressions for the one- and two-photon interference curves that we expect to measure, and analyzed the hybrid behavior presented in the two-photon joint probability. The complementarity relation for the visibilities of these interferences was also theoretically discussed and an experiment, based on a type II SPDC source of HES, was performed in order to verify this relation. Our experiment corresponds to $12$ tests for this complementarity relation.

As a direct application, we mention that the hybrid two-photon interference can be used to quantify the hybrid entanglement. In particular, it is possible to choose which type of measurement, the two-photon polarization curve or the two-photon Young's interference, is best suited for the experimental entanglement determination. Another possibility is the investigation of complementarity in higher dimensional quantum systems~\cite{Bergou2,Gao,Suter} by generating multi-qubit and qubit-qudit HESs. The generation of the latter class of states was already discussed in Ref.~\cite{Hibr} and requires the use of a birefringent multi-slit.

\begin{acknowledgments}

We would like to thank C. H. Monken for lending us the sanded quarter-wave plates used in the double slit. This work was supported by Grants Milenio ICM P06-067F, CONICYT~PFB08-024, PBCT~Red21, FONDECYT~11085055 and FONDECYT~11085057. S. P\'{a}dua acknowledges the support of CNPq, FAPEMIG and National Institute of Science and Technology in Quantum Information, Brazil. M. Santiba\~{n}ez thank to CONICYT for scholarship support.

\end{acknowledgments}


\begin{references}
\bibitem{Note} The same thing happens when the idler detector is fixed at $x_i=\frac{\pi z}{2kd}$, and the signal analyzer is rotated. There will be no interference at all.
\end{references}
\end{document}